\documentclass[prd,preprint,showpacs,superscriptaddress]{revtex4}

\usepackage{amsfonts,amssymb}
\def\trace{\,\textrm{Tr}\,}
\begin{document}
\def\new#1{#1}

\preprint{gr-qc/0411076}
\title{Classical limit of quantum gravity in an accelerating universe}

\author{Frederic P. Schuller}
\email{fschuller@perimeterinstitute.ca}
\affiliation{Perimeter Institute for Theoretical Physics, 31 Caroline Street N, Waterloo N2L 2Y5, Canada}
\author{Mattias N.R. Wohlfarth}
\email{mattias.wohlfarth@desy.de}
\affiliation{II. Institut f\"ur Theoretische Physik, Universit\"at Hamburg, Luruper Chaussee 149, 22761 Hamburg, Germany}

\begin{abstract}
A one-parameter deformation of Einstein-Hilbert gravity with an inverse Riemann curvature term is derived as the classical limit of quantum gravity compatible with an accelerating universe. This result is based on the investigation of semi-classical theories with sectional curvature bounds which are shown not to admit static spherically symmetric black holes if otherwise of phenomenological interest. We discuss the impact on the canonical quantization of gravity, and observe that worldsheet string theory is not affected.   
\end{abstract}
\maketitle

Einstein-Hilbert gravity is by birth a classical theory. Without additional provisions, it fails both at very short and very large distances. At short range, the generic occurrence of singularities \cite{HaPe} predicts the breakdown of spacetime itself. At long range, recent observations \cite{obs_acceluniv,obs_acceluniv2} indicate that the universe currently undergoes accelerating expansion, requiring the addition of a vacuum `dark' energy which is most straightforwardly modelled by a positive cosmological constant (or by scalar fields violating the strong energy condition, e.g. \cite{ToWo03}). If both the solution to the cosmological constant problem \cite{qftcosmo,qftcosmo2} and a resolution of spacetime singularities are to arise from a future quantum theory of gravity, the latter must generate corrections to Einstein-Hilbert gravity that modify both its long and short range behaviour.

In this Letter, we systematically investigate gravity theories whose solutions respect lower and upper sectional curvature bounds. Two-sided bounds can be motivated from quantum gravity heuristics, and allow to draw  farther-reaching conclusions than in the case of upper bounds only, which the authors studied in \cite{ScWo04}. In particular, we prove that static spherically symmetric solutions admitting a Kepler regime do neither admit a black hole singularity nor a horizon, thus featuring a short range behaviour radically different from Einstein-Hilbert gravity. Our approach deliberately makes only weak assumptions about the exact character of an underlying quantum spacetime structure, resulting in an ambiguity reflected in the fact that there are as many theories with two-sided curvature bounds as there are holomorphic functions on an annulus. Taking the classical limit by removal of both curvature bounds leads to a surprise: the resulting dynamics are unique and given by a one-parameter deformation of Einstein-Hilbert gravity with an inverse Riemann curvature term that vanishes only for a non-accelerating universe. This entirely unexpected dequantization result has significant  implications for the quantization of gravity. The deformed action contains additional degrees of freedom compared to standard general relativity, and these should be included in a canonical quantization. The case of spacetime dimension two, however, is special in that the deformation vanishes there identically. In particular, our findings leave the worldsheet action in string theory unaltered.

Before developing the above results in detail, we present heuristic arguments for the emergence of sectional curvature bounds from some form of quantum spacetime. First recall that while any two inertial observers on flat spacetime will
agree on the vacuum state of a quantum field, this no longer holds on a curved background \cite{BiDa}. Two nearby inertial, i.e., freely falling, observers generically set up different normal coordinates. Hence, even if one observer detects a quantum vacuum with respect to his coordinate system, the second observer will detect quantum excitations due to a non-trivial Bogoliubov transformation between the two coordinate systems. The excited state will be thermal in good approximation, if the relative acceleration is high compared to the inverse proper time during which the second observer operates his particle detector. This follows from the Unruh temperature for detectors with finite lifetime \cite{Rovelli}, which asymptotically equals the Unruh temperature \cite{Unruh} in the above-mentioned limit. The extension of the Unruh effect to detectors with finite lifetime is remarkable, because it implies that the Unruh effect holds quasi-locally, and thus for tidal accelerations by the strong principle of equivalence. (While it is difficult to make this precise for generic spacetimes, de Sitter space may be considered an instructive example. For an adiabatic vacuum of the quantum field, a comoving detector will measure a Gibbons-Hawking temperature proportional to the square root of the curvature \cite{GiHa}.) The second step in linking sectional curvature bounds to quantum gravity consists in Sakharov's observation that a minimal fundamental length scale implies an upper bound of the order $\sqrt{c^5\hbar/Gk_B^2}$ on the temperature of any thermal radiation \cite{Sakharov}. Sakharov's heuristic considerations employ the equation of state of thermal radiation at the extreme density of one quantum per Planck volume. Combining Sakharov's maximum temperature with the Unruh effect therefore suggests an upper bound of the order of the Planck scale on tidal accelerations, and thus on the sectional curvature.

Dual to the corresponding small length scale $\lambda$ is a large one which we denote by $\Lambda$. The required dimensionless hierarchy is generated by the discretization of a $d$-dimensional spacetime region of volume $V$ (in the spirit of Sakharov's construction) into $N=V/\lambda^{d}$ points. We will see later that sectional curvature bounded gravity possesses a solution of discretization-independent constant curvature $k = (\lambda\Lambda)^{-1}$ precisely when $\Lambda = N^{2/d} \lambda$. It is interesting to note that $k$ then also coincides with the prediction \cite{Sorkin2} of the cosmological constant in four dimensions from the causal set approach \cite{Sorkin} to quantum gravity.

We would like to emphasize that the above reasoning must remain a heuristic one in the absence of a complete quantum theory of gravity. The line of argument, however, presents one more example for the intimate relationship between the concepts of length and acceleration scales. Maximal covariant, rather than tidal, acceleration for particle motion has been discussed in different contexts by several authors \cite{Caianello,Brandt,Schuller1,Schuller2,Toller}.  A minimal acceleration principle in gravity is the basis for modified Newtonian dynamics \cite{Mond}.

We now turn to a detailed derivation of the results stated in the introduction. The geometrical object central to our investigation is the sectional curvature of a metric manifold,
\begin{equation}\label{secdef}
S(X,Y) = \frac{R(X,Y,X,Y)}{G(X,Y,X,Y)}\,,
\end{equation}
where $R$ denotes the Riemann-Christoffel tensor and $G(X,Y,Z,W)=g(X,Z)g(Y,W)-g(X,W)g(Y,Z)$, so that the denominator appearing above is the squared area of the parallelogram spanned by the vectors $X$ and $Y$. Physically, the sectional curvature corresponds to the frequency of the oscillation of a geodesic around a nearby geodesic with tangent $X$, connected to the first by a vector field $Y$ satisfying $[X,Y]=0$. It is easily checked that the sectional curvatures at a point only depend on the plane spanned by $X$ and $Y$, rather than on these two vectors individually. The sectional curvatures therefore provide a normalized measure for the tidal accelerations. Knowledge of the sectional curvatures for all possible planes determines the curvature tensor of a Riemannian or semi-Riemannian manifold. The physically relevant case of Lorentzian manifolds, however, imposes some restrictions to which we must attend carefully. In this case, sectional curvatures are only defined with respect to non-null planes, i.e., those planes spanned by vectors $X$ and $Y$ such that $G(X,Y,X,Y)\neq 0$. We will assume that the differentiable manifold in question has $d$ dimensions and Lorentzian signature $(-,+,\dots,+)$.

Motivated by the heuristic result arrived at in the introduction, namely that sectional curvatures are bounded by a quantum mechanism, we now aim at rigorously imposing a restriction of the form
\begin{equation}\label{secbound}
\Lambda^{-2} \leq |S(E)| \leq \lambda^{-2}
\end{equation}
in a covariant way, where $E$ denotes the set of non-null planes and $\lambda<\Lambda$ are two length scales. As discussed in \cite{ScWo04}, this program faces an immediate problem: for space-time dimension $d>2$, there exist rigidity theorems \cite{rigidity} stating that the only Lorentzian manifolds with everywhere bounded sectional curvatures are those of constant curvature. As this is clearly too restrictive for a viable gravity theory, we must find a covariant restriction of the set $E$ of all non-null planes to a subset $E'$, and impose (\ref{secbound}) only on that subset. The only alternative to such a restriction, studied by Andersson and Howard \cite{Andersson}, unfortunately does not allow to bound the absolute value of the sectional curvature.
 The impossibility of bounding all tidal accelerations of a Lorentzian manifold has an analogue in electrodynamics. The invariants of the electromagnetic field strength are given by $E\cdot B$ and $E^2 - B^2$. Clearly, one cannot impose a covariant bound on both the electric and magnetic fields, due to the minus sign in the second invariant, which is a consequence of the Lorentzian spacetime signature. However, the example of Born-Infeld electrodynamics \cite{BI} shows that despite the failure to bound the complete field strength, one may still be able to devise dynamics with regularized solutions. This is also true in the case of curvature bounds in gravity, as we will see.

The maximal subset of planes, on which one can impose sectional curvature bounds on Lorentzian manifolds without
running into the domain of the rigidity theorems, is given by the construction of \cite{ScWo04}, which we will outline
in the following. Due to the symmetries of the Riemann tensor (arising from a metric-compatible connection),
$R^{ab}{}_{cd}$ defines a linear map on the $d(d-1)/2$-dimensional space $\bigwedge^2$ of antisymmetric two-tensors.
Moreover, $R^{ab}{}_{cd}$ is symmetric with respect to the induced metric on that space, which is given by the tensor
$G$ introduced in (\ref{secdef}). It is easily verified that $G$ shares all the symmetries of the Riemann tensor. Over
a Riemannian manifold, $G$ is a positive definite bilinear form on $\bigwedge^2$; hence $R$ can be diagonalized with
real eigenvalues. For Lorentzian manifolds, the metric $G$ has the indefinite signature $(d-1,(d-1)(d-2)/2)$.
In this case, orthogonal diagonalizability with real eigenvalues is guaranteed if $R$ is a Pesonen operator \cite{Pesonen}, i.e., if $G(R(\Omega),\Omega) \neq 0$ for all $\Omega$ with $G(\Omega,\Omega)=0$.  
Whether fully diagonalizable or not, the eigenvectors of the Riemann endomorphism are of particular importance
for the construction of a restricted set $E'$ of planes on which we will impose the curvature bounds (\ref{secbound}).
We only discuss the non-trivial Lorentzian case. Let $\Omega_I \in \bigwedge^2$ be the maximally $d-1$ Riemann eigenvectors with
$G(\Omega_I,\Omega_I)<0$, and $\Omega_{\bar I} \in \bigwedge^2$ the maximally $(d-1)(d-2)/2$ eigenvectors with $G(\Omega_{\bar
I}, \Omega_{\bar I}) > 0$. Note that neither of the $\Omega_I, \Omega_{\bar I}$ themselves necessarily describe
planes; only antisymmetric two-tensors that can be written as an anti-symmetrized product of two vectors correspond to
planes. Such antisymmetric two-tensors are called simple, and present a polynomial subset of $\bigwedge^2$. We can now
characterize the restricted set of planes $E'$ as the simple elements lying in either the linear span of the
$\Omega_I$ or the linear span of the $\Omega_{\bar I}$. That $E'$ is the maximal set of planes to which one can
restrict the sectional curvature map in an algebraically sensible way, is explained in detail in \cite{ScWo04}. Evidently, the higher the degree to which the Riemann tensor is diagonalizable, the larger the set $E'$.

We are now prepared to state, in precise terms, a simple criterion for a manifold to feature sectional curvature bounds on the restricted set $E'$. The bounds (\ref{secbound}) with $E$ restricted to $E'$ are equivalent to the bounds
\begin{equation}\label{riembound}
\Lambda^{-2} \leq |e_R| \leq \lambda^{-2}
\end{equation}
on all eigenvalues $e_R$ of the Riemann tensor, regarded as an endomorphism $R^{[ab]}{}_{[cd]}$ on the space $\bigwedge^2$ of antisymmetric two-tensors. The inspection of maximally $d(d-1)/2$ eigenvalues is a very convenient criterion, as it circumvents the explicit construction of the set $E'$ and any non-tensorial relation of the form (\ref{secbound}). Condition (\ref{riembound}), in contrast, is manifestly covariant, as the characteristic polynomial for the Riemann tensor certainly is. We remark that over a Riemannian manifold, rather than a Lorentzian one, the restricted set of planes $E'$ coincides with the set of all planes $E$, as $G$ is positive definite in this case. We finally owe the reader proof that the sectional curvature bounds on the restricted set of planes $E'$ do indeed circumvent the rigidity theorems mentioned before. It is
sufficient to give an example: the four-dimensional spacetime
\begin{equation}
ds^2=-(1+r^2/\Lambda^2)^2dt^2+\frac{dr^2}{1+r^2/\Lambda^2}+r^2d\Omega^2\,,
\end{equation}
where $d\Omega^2$ is the line element on the unit two-sphere, satisfies both upper and lower curvature bounds with $\Lambda^{-2}\le |e_R|\le (\Lambda/2)^{-2}$, while its curvature is non-constant.

As an instructive application of the above criterion we discuss static spherically symmetric spacetimes with sectional
curvature bounds in four dimensions. Consider a spacetime ansatz of the form $ds^2 = -A(r) dt^2 + B(r) dr^2 + r^2
d\Omega^2$, where $A$ and $B$ are smooth on their respective domains. We assume non-degeneracy of the given metric,
which implies $AB>0$, as degeneracy corresponds to a breakdown of the Lorentzian signature. The Riemann tensor
$R^{[ab]}{}_{[cd]}$ is already diagonal in the basis $\{[tr],[t\theta],[t\phi],[r\theta],[r\phi],[\theta\phi]\}$ which
the standard Schwarzschild coordinates $\{t,r,\theta,\phi\}$ induce on the space $\bigwedge^2$, so that the
eigenvalues are easily obtained. The evaluation of the eigenvalue bounds results in conditions on the functions $A$
and $B$ and narrows the spectrum of admissible spacetimes of the above form to essentially two types. In type I we
have $0< B(r)\leq 1$ for all $r$. Then $A$ is positive; one can also show that $A$ is strictly monotonous and cannot
possess poles. For $A'>0$ we hence obtain a phenomenologically viable class of solutions, which notably possess
neither horizons nor curvature singularities. Type II cannot accommodate, without naked singularities, any spacetime
yielding an attractive gravitational field in some intermediate region between $\lambda < r < \Lambda$, as required
phenomenologically. This can be seen from the curvature bounds that imply $A$ is strictly monotonous and exclude
certain ranges of values. Attractive gravity in the intermediate region is only possible if $|A|\rightarrow\infty$
somewhere in this region. But then the metric density and hence the spacetime feature a singularity. We conclude that
any phenomenologically tenable gravity theory with sectional curvature bounds does not contain static spherically
symmetric spacetimes with singularities shielded by a horizon, i.e., static spherically symmetric black holes. The
regularization of the Schwarzschild singularity at this semi-classical level is in accordance with recent work on the       quantization of the Schwarzschild solution \cite{Winkler}, and general results on properties of singularity-free static spherically symmetric spacetimes \cite{Holdom}.

We now turn from purely kinematical considerations to the construction of gravitational dynamics whose solutions obey lower and upper sectional curvature bounds.
Most stringently, the desired inequalities may be enforced by equations of motion containing a power series converging precisely on the domain (\ref{riembound}) allowed by the curvature bounds \cite{kempf}. (There might be other possibilities; for example, one might choose equations of motion which become singular at the boundaries of the allowed domain. Although solutions in this case might not be able to cross the singularities, they would fall into two classes, one obeying the desired bounds, the other one obeying the logical opposite.)
We choose to generate the power series in the equations of motion from an action as follows.
Let $\lambda < \Lambda$, and consider a holomorphic function $f$ with branch cuts along the real intervals $(-\infty,-\lambda^{-2})$, $(-\Lambda^{-2},\Lambda^{-2})$, and $(\lambda^{-2},\infty)$. Then $f(z)$ has a Laurent series expansion that converges absolutely on the annulus $\Lambda^{-2}<|z|<\lambda^{-2}$ and possibly points of its boundary, but nowhere else. As we want to devise non-strict bounds, it is advantageous to express $f$ in terms of the dimensionless coefficients $a_{-n}$ and $a_n$ of two Taylor series $\sum_{n=1}^\infty a_{-n} x^n$ and $\sum_{n=1}^\infty a_n x^n$ which are both absolutely convergent for $|x|\leq1$. Then we consider functions $f$ with Laurent series
\begin{equation}\label{eq.expansion}
f(z) = \sum_{n=0}^\infty\left(a_{-n} \Lambda^{-2n-2} z^{-n} + a_n \lambda^{2n-2} z^n\right),
\end{equation}
which by construction converge absolutely on $\Lambda^{-2} \leq |z| \leq \lambda^{-2}$. We now stipulate the action
\begin{equation}\label{eq.action}
S = \int_M \sqrt{-g} \trace f(R)\,,
\end{equation}
where $R$ is the Riemann tensor regarded as a linear map on the space of antisymmetric two-tensors, and the trace is defined as ${\trace f(R) = f(R)^{[ab]}{}_{[ab]}}/2$. Diffeomorphism invariance implies the Noether constraint $\nabla_i (\delta S/\delta g_{ij}) = 0$, and hence matter can be coupled in standard fashion, simply by adding an appropriate matter action $S_M$ to $S$.

The above action contains inverse powers of the Riemann tensor, which means it assumes that the Riemann tensor as a map on $\bigwedge^2TM$ is invertible. We will see below that this assumption translates into a condition on the space of solutions to the equations of motion. It follows that this space can only contain spacetimes whose Riemann tensor is invertible. Thus specific classes of spacetimes $M$ can never solve, e.g., unwarped topological products $M=M_p\times M_q$ of $p,q$-dimensional spaces, for which the Riemann tensor has at least $pq$ vanishing eigenvalues.
Regarding the method of variation, by which to obtain equations of motion from the action (\ref{eq.action}), consider the following point. The definition of sectional curvature depends on the connection being metric compatible; otherwise the Riemann tensor cannot be fully reconstructed, nor is the sectional curvature well-defined on the space of 2-planes. Hence a Palatini procedure, where the metric $g$ and an affine connection $\Gamma$ are varied independently, seems unnatural in the context of our construction. The equations of motion are therefore derived from the total action by variation with respect to the spacetime metric $g$. This is only feasible if the expansion of $f$ can be re-ordered, hence the restriction to absolute convergence, and hence holomorphicity. The gravitational field equations read
\begin{equation}\label{eq.motion}
\frac{1}{2} f'(R)^{cd(i}{}_{b} R^{j)b}{}_{cd} - g^{ij} \trace f(R) - \nabla_b \nabla_c f'(R)^{c(ij)b} = T^{ij}\,,
\end{equation}
where $T^{ij}$ is the energy momentum tensor of $S_M$, and the sign convention $R^a{}_{bcd} = \partial_c\Gamma^a_{bd} + \Gamma^a{}_{ec}\Gamma^e{}_{bd} - (c \leftrightarrow d)$ has been used. As explained above, the appearance of $\trace f(R)$ in the equations of motion and the convergence properties of the function $f$ guarantee that any solution satisfies sectional curvature bounds. Also, any such solution will have an invertible Riemann tensor.

Removal of the lower and upper curvature bounds corresponds to taking the classical limit. Without further restrictions, however, there is no unique way how to take the two limits $\lambda\rightarrow 0$ and $\Lambda\rightarrow\infty$ with respect to each other. A physically motivated prescription for taking these limits in a controlled fashion lies in the choice of vacuum solutions of the desired classical theory, as we will now show. For finite $\Lambda$, the curvature bounds (\ref{riembound}) exclude Minkowski space. But for a number of purposes, such as stability checks, perturbative solutions or cosmological phenomenology, it is advantageous to have some space of constant curvature $\pm k$ as an exact vacuum solution.
The adoption of a non-flat vacuum is indeed a natural choice given the recent observational evidence for an accelerating universe. Note that the introduction of a vacuum curvature $k$ does not introduce a third independent length scale; in fact, the curvature $k$ is defined by a length scale $\Lambda$ relative to an a priori arbitrary power of the scale $\lambda$, according to $k=\lambda^{2\alpha}\Lambda^{-2\alpha-2}$.
The value of $\alpha$ is actually fixed by the requirement that the equations of motion (\ref{eq.motion}) are then indeed solved for the (anti-)de Sitter spaces $R^A{}_B = \pm k \delta^A_B$ in vacuo, which translates into the condition
\begin{equation}\label{eq.exactcondition}
\sum_{n=0}^\infty \left(a_{-n}\frac{2n+d}{\Lambda^{2n+2}}(\pm k)^{-n} -
a_{n}\frac{2n-d}{\lambda^{2-2n}}(\pm k)^{n}\right) = 0\,.
\end{equation}
There are of course many ways to define coefficients $a_n$ in order to satisfy this condition. However, for the exact solutions with curvature $\pm k$ to be of use in perturbation theory (which was one of the motivations for the present construction), we demand that (\ref{eq.exactcondition}) vanishes order by order. This uniquely determines the $a_{-n}$ in terms of the $a_n$, using our expression for $k$,
\begin{equation}\label{eq.predef} a_{-n} =
\frac{2n-d}{2n+d} \left(\frac{\lambda}{\Lambda}\right)^{2n(1+2\alpha)-2}a_n\,,\qquad n>0\,,
\end{equation}
and $a_0=0$. Before accepting this definition, we must check whether $\sum_{n>0} a_{-n} x^n$ converges absolutely on $|x|\leq 1$ (otherwise, we would not enforce sectional curvature bounds) if and only if $\sum_{n>0} a_n x^n$ does (which is guaranteed by the choice of the $a_n$). This is the case if and only if the sequence $a_{-n}/a_n$ converges to a non-vanishing constant as $n \rightarrow\infty$. From (\ref{eq.predef}) it is obvious that this is the case only for $\alpha=-1/2$, determining $ k = (\lambda\Lambda)^{-1}$.
The curvature scale $k$ of the exact de Sitter and anti-de Sitter solutions hence emerges as the geometric mean of the upper and lower curvature bounds, so that perturbation theory around these exact solutions is a sensible undertaking.
We may think of the scale $\lambda$ as inducing a discretization $N=V/\lambda^d$ of a given spacetime volume $V$ into $N$ points. Then, in the classical limit, the gravitational dynamics should be discretization-independent. In particular, $k$ should not depend on $N$, which requirement determines a hierarchy $\Lambda=N^{2/d}\lambda$, and $k=V^{-2/d}$. This argument shows that the classical limit $N\rightarrow\infty$ takes $\lambda\rightarrow 0$ and $\Lambda\rightarrow\infty$ while keeping the volume $V$ of the classical spacetime region, and thus $k$, constant. This singles out the unique classical action
\begin{equation}\label{inverseR}
S_\textrm{class.} = \int_M \sqrt{-g} \,a_1 \trace\!\left( R -
\frac{d-2}{d+2} k^2\, R^{-1} \right)
\end{equation}
from the vast class (\ref{eq.action}). For $k=0$ we reassuringly obtain Einstein-Hilbert gravity. Remarkably, the same holds for $d=2$, so that the string worldsheet action is not affected.

For non-vanishing vacuum curvature $k$ in $d>2$ the classical limit is distinctly different from theories whose Lagrangian is a function $f$ of the Ricci scalar because generically $\trace (R^{-1})$ cannot be expressed as such a function.
The latter type of theories have attracted a lot of attention recently, as phenomenological models explaining the present cosmic acceleration \cite{invR2,KrAl04}. The origin of inverse curvature corrections in particular has been argued to arise from time-dependent M-theory compactifications \cite{Odintsov}.
Theories depending only on the Ricci scalar allow for a reformulation as scalar-tensor theories, where the effects of curvature corrections are absorbed into matter fields \cite{scaten}. Exactly this reducibility, however, lies at the heart of proofs demonstrating the inconsistency of scalar $1/R$ gravity theory with observations in the solar system \cite{Chiba} or particle physics \cite{Flanagan}. (Further potentially problematic aspects of these theories are discussed in \cite{invR3}.)
These arguments do not extend to the classical theory (\ref{inverseR}), where the additional dynamical degrees of freedom cannot be made explicit by any transformation of the metric, due to the disparate dimensionality of the space of metrics and the space of Riemann tensors.
For a static spherically symmetric ansatz, the geometrical origin of the theory as a limit of curvature bounded theories further turns out to determine six out of seven boundary conditions required to solve the higher-derivative equations of motion, leaving only the mass as a free integration constant. These observations alone indicate that the theory (\ref{inverseR}) does not suffer from persistent problems of other, more ad hoc, modifications of Einstein-Hilbert gravity.

In summary, by removing quantum gravity motivated curvature bounds in a controlled way, we have derived a novel classical gravity theory that is not immediately invalidated by standard inconsistency arguments, and which sheds a new light on the debate of whether pure gravity can be quantized:
the generic appearance of additional degrees of freedom, in more than two spacetime dimensions, indicates that a canonical quantization of Einstein-Hilbert gravity apparently misses out on some degrees of freedom, while worldsheet string theory is unaffected.

The authors wish to thank Cedric Deffayet, Laurent Freidel, Achim Kempf, Mark Trodden and Damien Easson for stimulating discussions.

\end{document}